\documentclass[
 reprint,
 amsmath,amssymb,
 aps,
 prl,
 superscriptaddress,
]{revtex4-2}

\usepackage{graphicx}
\usepackage{dcolumn}
\usepackage{bm}
\usepackage{float}
\usepackage{listings}
\usepackage{xcolor}
\usepackage{braket}
\usepackage{amsmath}
\usepackage{MnSymbol}
\usepackage{svg}
\usepackage{textcomp}
\usepackage{mathtools}
\usepackage[normalem]{ulem}
\usepackage{hyperref}
\usepackage{silence}
\usepackage{array}
\usepackage[english]{babel}

\usepackage{xr}

\newcolumntype{P}[1]{>{\centering\arraybackslash}p{#1}}
\newcolumntype{M}[1]{>{\centering\arraybackslash}m{#1}}

\begin{document}
\title{Electronic primary thermometry --- experimental comparison of the Coulomb Blockade and Shot Noise Thermometer}

\author{Eemil Praks}
\affiliation{Pico group, QTF Centre of Excellence, Department of Applied Physics, Aalto University School of Science}
\author{Libin Wang}
\affiliation{Pico group, QTF Centre of Excellence, Department of Applied Physics, Aalto University School of Science}
\author{Yu-Cheng Chang}
\affiliation{Pico group, QTF Centre of Excellence, Department of Applied Physics, Aalto University School of Science}
\author{Renan P. Loreto}
\affiliation{VTT Technical Research Centre of Finland Ltd, P.O. Box 1000, FI-02044 VTT Espoo, Finland}
\author{Mika Prunnila}
\affiliation{VTT Technical Research Centre of Finland Ltd, P.O. Box 1000, FI-02044 VTT Espoo, Finland}
\author{Joonas T. Peltonen}
\affiliation{Pico group, QTF Centre of Excellence, Department of Applied Physics, Aalto University School of Science}
\author{Jukka P. Pekola}
\affiliation{Pico group, QTF Centre of Excellence, Department of Applied Physics, Aalto University School of Science}

\date{\today}

\begin{abstract}
Since the redefinition of the kelvin in 2019, new methods of primary thermometry have been considered to replace the currently agreed temperature scale in the very high and low temperature limits. These new methods should provide improved uncertainties and, most importantly, a more direct link to the definition of the kelvin. 
We present an experimental comparison of two such primary thermometers working in the mK region: the Coulomb Blockade Thermometer (CBT) and the Shot Noise Thermometer (SNT). Both thermometers measure temperature hinging only on the natural constants $k_\mathrm{B}$, and $e$. 
Furthermore both of them are based on electron tunneling current and, thereby,  need only electrical measurements, enhancing the practicality.
CBT and the SNT are inter-compared in a range of 20 mK to 235 mK. The results show that the agreement of SNT and CBT is approximately within 2.5~\% in this range. Basic measurement uncertainty is analyzed and we show that uncertainty in the measurement frequency can cause significant error to temperature measurement of the SNT at low temperatures where finite frequency plays a role.
\end{abstract}

\maketitle

Temperature has immense influence on many fundamental processes across different domains of science. 
In 2019, the SI unit of temperature, the kelvin, was redefined on the fixed Boltzmann constant $k_\mathrm{B}$~\cite{noauthor_international_2019}. Despite this, few of the commonly used thermometers can trace their measurements directly to the constant, relying instead on the older definition based on the triple point of water.
From the metrological perspective, measurement of temperature is also generally difficult and imprecise compared to that of other quantities.
As an example, atomic clocks achieve relative instabilities below $10^{-16}$~\cite{guena_progress_2012} and Josephson voltage standards have relative uncertainties below $10^{-8}$~\cite{kohlmann_josephson_2003}. By comparison, the relative uncertainty in temperature measurements is on the order of $10^{-5}$~\cite{neuer_comparison_2001, rourke_refractive-index_2019}. At low temperatures, below ten kelvin, relative uncertainty is on the order of $10^{-2}$~\cite{gao_primary_2026}.
Despite the reduced precision, temperature remains a very important parameter for multiple experiments in the mK range.
For instance, qubit decoherence~\cite{catelani_decoherence_2012}, phase transitions~\cite{vojta_quantum_2003}, fractional quantum hall effect~\cite{dagosta_temperature_2005}, all strongly depend on temperature. 

Currently, the most common low temperature thermometers are based on the temperature dependent resistivity of oxide compounds such as zirconium oxide and ruthenium oxide (RuOx). Both examples are commercially available from multiple sources. Unfortunately the calibration accuracy is poor, usually $\geq 10\ \%$ at 10 mK and below~\cite{noauthor_lakeshore_nodate}. A higher accuracy alternative is the cerium magnesium nitrate (CMN) thermometer, which relies on the temperature dependent magnetic susceptibility of CMN~\cite{greywall_fast_1989}. It provides superior accuracy, at the cost of reduced practicality and increased sensitivity to magnetic fields. Notably, all of the mentioned thermometers are secondary in character and thus rely on a calibration. This means that a thermometer is only as accurate as its calibration, and the precision degrades over time with thermal cycles. 

These drawbacks are the main motivation for primary thermometers, that is, thermometers based on a well understood physical system from which temperature can be determined without a known reference. Primary thermometers, by definition, eliminate the problems of calibration and long term drift.
The most prominently used primary thermometer at mK temperatures is the Johnson noise thermometer (JNT)~\cite{engert_noise_2012}. Other well known techniques include acoustic gas thermometry (AGT)~\cite{moldover_acoustic_2014}, which is based on the absolute temperature dependence of the speed of sound in an ideal gas.
One of the greatest challenges with primary thermometers and precise thermometry in the mK-regime in general, is practicality. The devices tend to be difficult to operate and sensitive to external conditions, such as magnetic fields or vibrations.

Besides improved precision and accuracy, another motivation for the development of new primary thermometers is traceability.
As more precise measurements have been realized, the definitions of units are now based on natural constants~\cite{bureau_international_des_poids_et_mesures_systeme_2006, noauthor_international_2019}. Many new measurements can be relatively directly traced to these constants.
In this spirit, the kelvin was redefined in 2019~\cite{noauthor_international_2019} based on the fixed value $1.380649 \times 10^{-23} \ \mathrm{JK^{-1}}$ of the Boltzmann constant $k_\mathrm{B}$. 
The previous temperature scale, ITS-90~\cite{noauthor_internatioanl_1989} and its low temperature extension PLTS-2000~\cite{noauthor_provisional_2000} 
are still ultimately based on the old standard: triple point of water $T_\mathrm{TPW}$. The scales are expected to remain in use, but they are considered less reliable further away from $T_\mathrm{TPW}$.
In particular, the standards have poor agreement in their overlapping region between 650 mK and 1 K~\cite{pan_direct_2021}. For these reasons, new methods based on $k_\mathrm{B}$ are expected to become more relevant far from $T_\mathrm{TPW}$~\cite{fellmuth_kelvin_2016}.
Currently, the 2nd appendix of the SI brochure~\cite{noauthor_international_2019} defines a few primary thermometers, or practical realizations of the kelvin --- \textit{mise en pratique kelvin (MeP-K)}, most prominently AGT and JNT. The list of methods is expected to expand in the future. 

In this work, we present an experimental comparison of two primary thermometers utilizing different effects: the Coulomb Blockade Thermometer (CBT)~\cite{pekola_thermometry_1994} and the Shot Noise Thermometer (SNT)~\cite{spietz_primary_2003, sivre_electronic_2019}. We compare their performance and investigate their practicality. The thermometers are chosen for their true primary nature and fully electronic measurement; no external calibration constants are required beyond $k_\mathrm{B}$ and elementary charge $e$. This is contrary to JNT, where a single reference temperature is necessary, making it a relative primary thermometer. 
Another important advantage is the fully electronic measurement, unlike in AGT which relies on the assumption of an ideal gas. Electronic measurements are relatively simple to realize to a high accuracy and current and voltage already have standards defined on natural constants. 

\begin{figure}
    \centering
    \includegraphics[scale = 0.75]{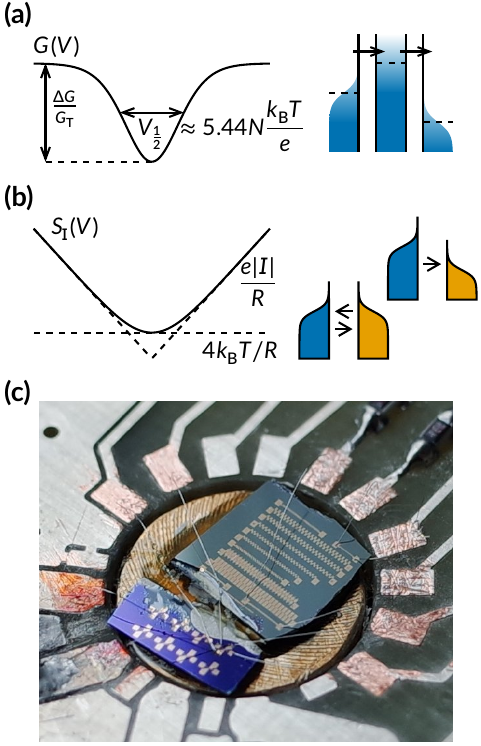}
    \caption{Schematic illustration of the operating principles of (a) CBT and (b) SNT. In CBT Coulomb barrier needs to be overcome by thermal fluctuations, whereas in SNT temperature determines the cross-over between thermal noise and Shot noise. (c) CBT and SNT are mounted to the same sample holder to minimize the possibility of thermal gradients between the two thermometers.}
    \label{fig:snt-cbt-schematic}
\end{figure}
CBT, first proposed and demonstrated in 1994~\cite{pekola_thermometry_1994}, consists of a chain of normal metal-insulator-normal metal (NIN) tunnel junctions. 
Due to Coulomb blockade, a tunnel junction exhibits a reduction in conductance near zero bias. In the limit of weak charging effects, the bias dependence of the normalized conductance $G/G_\mathrm{T}$ is dependent only on the junction parameters and temperature. Remarkably, the full voltage width at half maximum as shown in Fig.~\ref{fig:snt-cbt-schematic}(a), is fully determined by the temperature and natural constants as
\begin{equation}
    V_\frac{1}{2} \approx 5.44 N \frac{k_\mathrm{B} T}{e}.
\end{equation}
The result is completely independent of the charging energy $E_c$ and other system parameters. Conceptually, we are measuring the thermal occupation of electron states. Filled states follow the Fermi-distribution, and the density of state (DOS) on either side of a junction is constant to a high accuracy. By changing the potential difference, the Fermi levels are shifted with respect to each other, in turn changing the differential conductance. 

SNT utilizes a different property of tunnel junctions for thermometry: the discreteness of current carriers. First proposed in 2003~\cite{spietz_primary_2003}, the operating principle is remarkably simple. When no bias is applied, a junction exhibits temperature dependent current (or equivalently voltage) noise known as Johnson-Nyquist noise~\cite{nyquist_thermal_1928}. Bias voltage across a junction induces shot noise component, due to discrete tunneling events~\cite{schottky_uber_1918}. This depends only on the magnitude of the bias voltage. There is a gradual transition between these two noise regimes and the symmetric current noise is
\begin{widetext}
\begin{equation}
    S_I = \frac{1}{R} \left[ (eV+hf) \coth \left(\frac{eV + hf}{2 k_\mathrm{B} T}\right) + (eV-hf) \coth \left(\frac{eV - hf}{2 k_\mathrm{B} T}\right) \right].
    \label{eq:snt-noise-power}
\end{equation}
\end{widetext}
Here $f$ is the measurement frequency and $R$ is the junction resistance. 
The shape is depicted in Fig.~\ref{fig:snt-cbt-schematic}(b). 
When measured, the entire signal will be multiplied by the gain on the amplifier system, which may now always be known to a high precision. However, due to the temperature dependence of Johnson-Nyquist noise regime, and temperature independence of shot noise regime, temperature and amplifier gain can be independently determined.
Equation~\eqref{eq:snt-noise-power} also contains a quantum correction to this noise signal which becomes important when $f\gtrsim k_\mathrm{B} T / h$~\cite{spietz_shot_2006}. This correction is most apparent near zero bias and is particularly important in our measurements at the lowest temperatures.

To compare these two thermometers, measurements are performed in a Bluefors LD250 dilution refrigerator. Temperature differences are minimized by placing both thermometers on the same sample holder and measuring them in succession. The thermometer chips are shown in Fig.~\ref{fig:snt-cbt-schematic}(c). The cryostat temperature is positioned at values that are stabilized using its built-in PID control during the measurement. The PID loop uses a RuOx thermometer on the mixing chamber plate. 

\begin{figure}
    \centering
    \includegraphics[scale = 0.75]{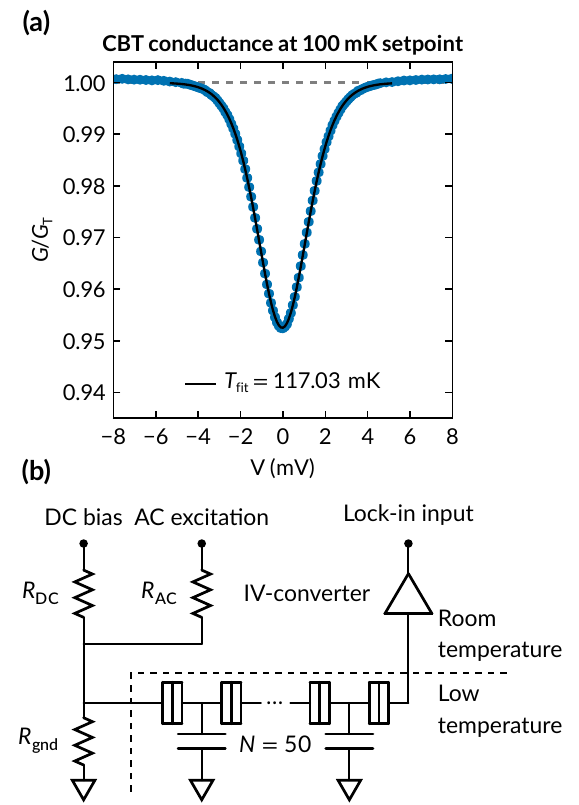}
    \caption{
    (a) Measured CBT conductance as a function of bias voltage $V$ and the best fit to it at 100 mK cryostat setpoint. 
    (b) Schematic representation of the CBT measurement setup. }
    \label{fig:cbt-setup-schematic}
\end{figure}
The measured CBT is a $N = 50$ junction array fabricated with Dolan bridge double angle technique from manganese doped aluminum with aluminum oxide as the junction barrier. This provides a normal metal junction even in the absence of magnetic field~\cite{ruggiero_dilute_2004}.  There are also 10 $\mathrm{\mu m}$ thick copper cooling pads deposited by means of electroplating on each island between junctions. Copper islands have proven to be an effective method of thermalization to remove or reduce self heating~\cite{bradley_nanoelectronic_2016, samani_microkelvin_2022, sarsby_500_2020} 
, which can in principle be theoretically compensated for, but it introduces uncertainty into the measurement. 

Thermometry is performed using a standard lock-in measurement, depicted in Fig.~\ref{fig:cbt-setup-schematic}(b). The array is voltage biased with a low frequency AC signal with a DC offset and the resulting current is measured. The DC component is varied to perform the voltage sweep, and the AC component is used for the conductance measurement. 
In our experiment a voltage divider at room temperature is used to combine and scale down a bias voltage and the AC excitation signal from a Zurich Instruments MFLI lock-in amplifier. The return current is amplified by a Basel Precision Instruments trans-impedance amplifier and routed to the lock-in amplifier input for measurement. 

\begin{figure}
    \centering
    \includegraphics[scale = 0.75]{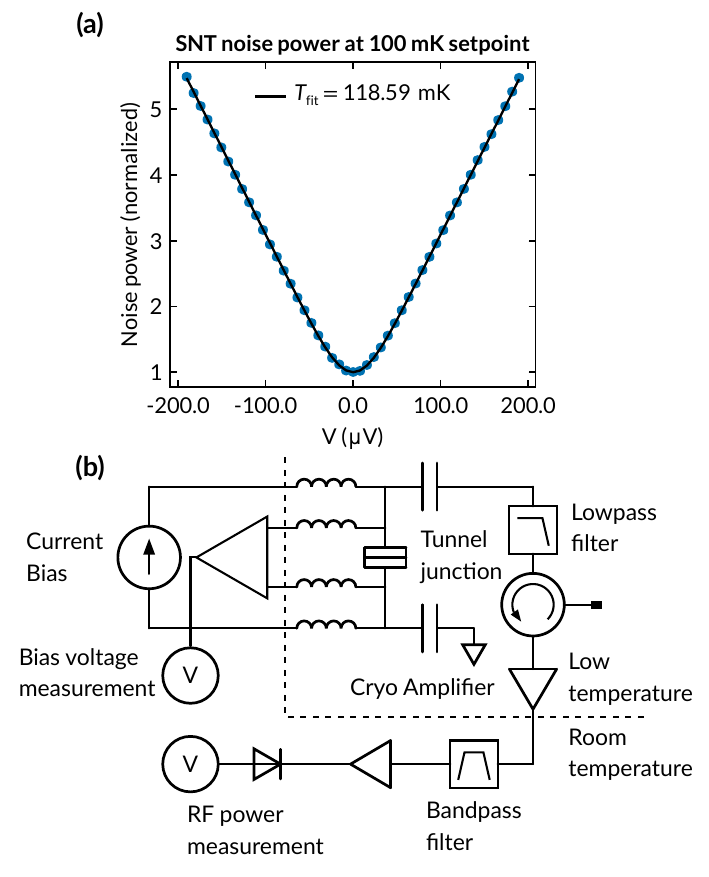}
    \caption{
    (a) Measured SNT noise power and best fit to it at 100 mK cryostat setpoint. 
    (b) Schematic of the SNT measurement setup. The DC biasing and RF components are separated using large inductors and capacitors mounted on the sample holder. 
    Final measurement bandwidth is fully determined by the components on the RF chain due to the wide bandwidth diode in the measurement. Noise outside the bandwidth does not affect the measurement. }
    \label{fig:snt-setup-schematic}
\end{figure}

Fabrication of the SNT junction is similar to that of the CBT array. The main difference is that the junction area is significantly larger to produce a junction resistance close to 50 $\mathrm{\Omega}$. 
By comparison, the CBT junctions have a resistance of about $3.5 \ \mathrm{k\Omega}$.
The low resistance of the SNT is important to improve the coupling between the junction and the RF chain used for the measurement. A copper layer is also present, although it is much thinner at approximately 400 nm and its purpose is to minimize voltage drops and thermal resistance in the device, which is essential to suppress Joule heating.
Both sides of the SNT junction are contacted by gold bonding wire to combat the bias dependent overheating. The standard Al bonding wires would turn superconducting, which would trap heat on the device. In operation, the voltage bias will have to be on the order of $k_\mathrm{B} T/e$, which combined with the $50 \ \Omega$ junction resistance results in a large current and pW level heat dissipation. Without efficient heat evacuation via the bonding wires, this heat is left to be evacuated mostly via electron-phonon channel~\cite{meschke_electron_2004}, which is typically insufficient at lower temperatures despite the relatively large bonding pads. 

The configuration of the SNT measurement follows the same principles as in~\cite{spietz_primary_2003, spietz_shot_2006}, see Fig.~\ref{fig:snt-setup-schematic}(b). The junction is current biased in a four-probe voltage measurement while RF noise is measured. DC and AC components are separated at the sample holder using large inductors and capacitors - 390 nH and 1 $\mathrm{\mu F}$ respectively. Most of the complexity is needed due to the very small signal to be measured. To prevent high frequency noise heating the junction, the DC lines are ThermoCoax\texttrademark\ lines which act as a low-pass filter. For the RF path, two circulators are used in combination with a VLFX-780 low-pass filter. 
The final bandwidth of the noise measurement is set by a multi-stage bandpass filter at room temperature with a central frequency around 750~MHz. The RF power measurement is done after further amplification using a diode type RF power meter.

Measurements and least squares fits are presented in Figs.~\ref{fig:cbt-setup-schematic}(a) and~\ref{fig:snt-setup-schematic}(a) for the two thermometers at $T\approx100 \ \mathrm{mK}$. Measurements at different temperatures can be found in supplemental material~\cite{supplement}. 
In the case of the CBT, the model used for least squares fit is based on a numerical solution to the master equation as described in~\cite{hirvi_one_1997}. 
In addition to the temperature and charging energy, the  fit parameters include a voltage offset and the normal-conductance value. Self-heating effects modeled in~\cite{meschke_electron_2004} are not considered in the fitting process. This effect is insignificant given the large volume of the electroplated device. 

Fitting model for the SNT is based on Eq.~\eqref{eq:snt-noise-power}, with a fixed frequency. The free parameters for the fit are the temperature and the gain of the system. 
We emphasize that there are no calibration parameters for either thermometer. 

\begin{figure}
    \centering
    \includegraphics[scale = 0.75]{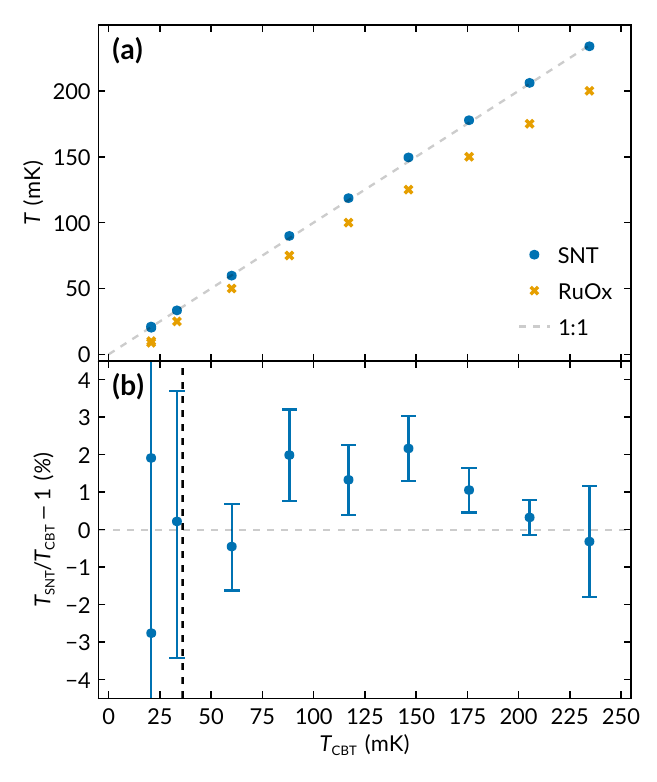}
    \caption{
    (a) Measured SNT and fridge RuOx thermometer temperatures as a function of CBT temperature. The mixing chamber plate RuOx temperatures deviate systemically from both the CBT and SNT.
    (b) Relative difference between the CBT and SNT. 
    The point where $k_\mathrm{B} T / h = 750 \ \mathrm{MHz}$ is marked by a dashed vertical line. Below this point, the uncertainty of the SNT measurement increases significantly. At the lowest temperature, slightly above 20 mK, the uncertainty is approximately $\pm 15 \ \%$. 
    }
    \label{fig:SNT-CBT-relative-difference}
\end{figure}

The central comparison of the two thermometers is presented in Fig.~\ref{fig:SNT-CBT-relative-difference}. The main result is the agreement of SNT and CBT within approximately 2.5~\% in a range from 20~mK to 235~mK. The readings of the refrigerator RuOx thermometer are also marked in Fig.~\ref{fig:SNT-CBT-relative-difference}(a). They deviate significantly from readings of the CBT and SNT. Despite factory calibration, we have found RuOx thermometers to be inaccurate with deviation between different RuOx sensors. More information is available in the supplemental material~\cite{supplement}.

Two main sources of uncertainty are estimated. First, standard errors based on the fit covariance are calculated for both thermometers. These are less than 0.5~\% for the SNT and less than 0.35~\% for the CBT.
By far the largest source of uncertainty is the precision of the SNT measurement bandwidth. The true noise signal is an integral over the measurement bandwidth, whereas the fit is made by assuming a single frequency. Numerical simulations indicate the mean frequency to provide relatively accurate thermometry, but the relation is not exact. 
Thus, additional error is calculated, based on an upper and lower bound to the fit frequency. These are taken to be $\pm 50 \ \mathrm{MHz}$ of the central frequency of 750 MHz. At the lowest temperature slightly above 20 mK the uncertainty is approximately $\pm 15 \ \%$. 
At temperatures $\geq h f/k_\mathrm{B}$, its contribution is negligible. In future measurements, the error can be mitigated by choosing a lower measurement frequency. 

The comparison demonstrates two fully independent primary thermometers in good agreement over a decade of temperature. This result is natural as both the CBT and SNT trace their measurements directly to $k_\mathrm{B}$. The small deviation is a strong indication that self heating effects common in on-chip devices can indeed be eliminated with careful design. It is also an additional proof of the correctness of the quantum frequency correction for the SNT~\cite{spietz_shot_2006}.

Both thermometers enable fully electronic measurement and a comparatively simple setup. Particularly in the case of CBT, the required circuitry and instrumentation are quite minimal. The required setup is more complicated for the SNT, but either thermometer could be realized in a well equipped calibration laboratory. Still, we think only the CBT has strong potential for general cryogenic thermometry, as it does not require low temperature RF components such as circulators. 

The main limitation of our work is the relatively narrow temperature range. The range of 20 mK to 235 mK is common for many quantum experiments, but both thermometers should be capable of far higher dynamic range. Particularly, the SNT should be capable of measurements up to room temperature. The dynamic range of a single CBT sensor is limited to approximately two orders of magnitude due to the requirement for weak but observable Coulomb blockade. We note that in our experiment the lowest temperature of approximately 20 mK is limited by the thermalization of the sample holder and could thus be extended by improving the sample holder design and mounting. 

Another important improvement is to reduce the uncertainty of the SNT measurement. The primary source of this is the poorly defined measurement bandwidth. Improving this is relatively straightforward through directly measuring the bandwidth of the total amplifier and filtering chain. Another way of reducing the uncertainty, as stated above, is to lower the measurement frequency. Previous realizations of the SNT~\cite{spietz_primary_2003,spietz_shot_2006} have chosen lower measurement frequencies. The choice depends primarily on practical factors, such as component availability, size constraints, and other setup parameters. Due to the close mounting it is unlikely that the remaining difference is caused by a thermal gradient. However, it would be possible to fabricate both thermometers on the same chip to further reduce the possibility of thermal gradients as in~\cite{iftikhar_primary_2016}. 

In this work we have considered only a single realization per thermometer and have not analyzed device to device variability. Due to the primary nature of CBT and SNT this variability is less interesting. Regardless, this is an important metric to verify, especially if either of the thermometers is to become a practical realization of kelvin (MeP-K). Both have interesting potential advantages for this purpose over current implementations, in particular a more direct link to $k_\mathrm{B}$, and lower implementation complexity. 

To conclude we have demonstrated consistency of two fully independent primary thermometers of about 2.5 \% from 20 mK to 235 mK. Practicality and implementation details of both thermometers have been discussed as well as the resulting uncertainties. The results are an important step in the evaluation of potential primary thermometers for a new 
realization of the kelvin and for general cryogenic thermometry.

\begin{acknowledgments}
We thank J. Luomahaara for useful discussions.

The research was financially supported by Business Finland project 5586/31/2024 and Research Council of Finland through the Finnish Quantum Flagship project (358877, Aalto University). R.L. and M.P. also ackowledge the European Union’s Horizon RIA and EIC programmes under Grants No. 101113086 SoCool and No. 101113983 Qu-Pilot and Research Council of Finland through the QTF Centre of Excellence (project No. 336817) and QMAT Centre of Excellence (project No. 374172). 

\end{acknowledgments}

\bibliography{references}

\clearpage
\section*{Supplementary information}

\begin{figure}[H]
    \centering
    \includegraphics[scale=0.75]{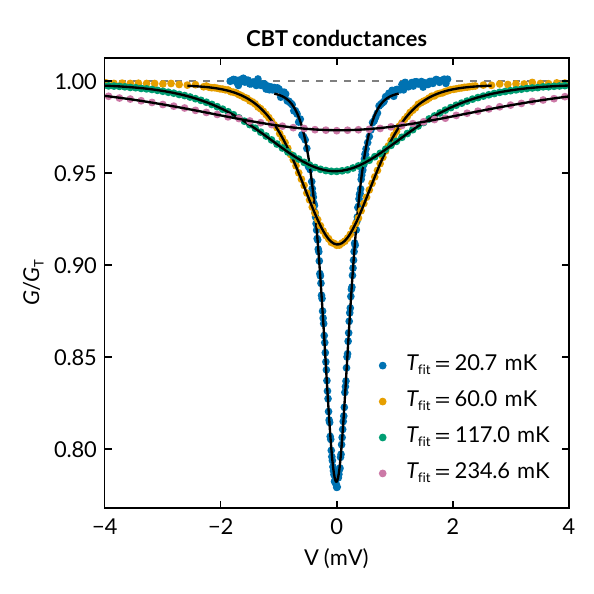}
    \caption{Measured CBT curves and least squares fits at a variety of temperatures. While the noise of the measurement increases at lower temperatures due to smaller excitation voltage, the overall uncertainty remains approximately constant due to the deeper conductance dip. }
    \label{fig:CBT-multi-T}
\end{figure}

\begin{figure}[H]
    \centering
    \includegraphics[scale = 0.75]{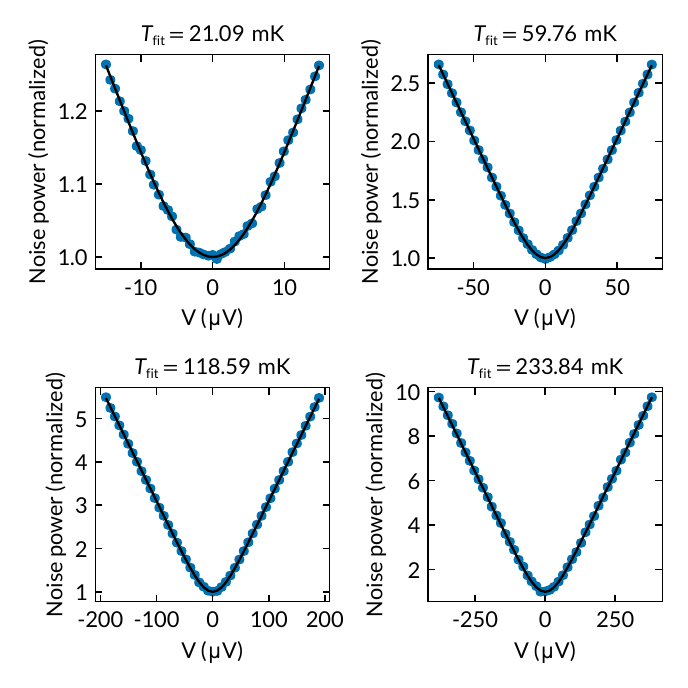}
    \caption{Measured SNT curves and least squares fits at a variety of temperatures. Note that the relative noise decreases when temperature increases as the signal gets larger. Another important observation is that at the lowest temperature, the curve exhibits a flatter bottom part. This is due to the quantum frequency correction increasing zero bias noise. }
    \label{fig:SNT-multi-T}
\end{figure}

\begin{figure}[H]
    \centering
    \includegraphics[scale = 0.75]{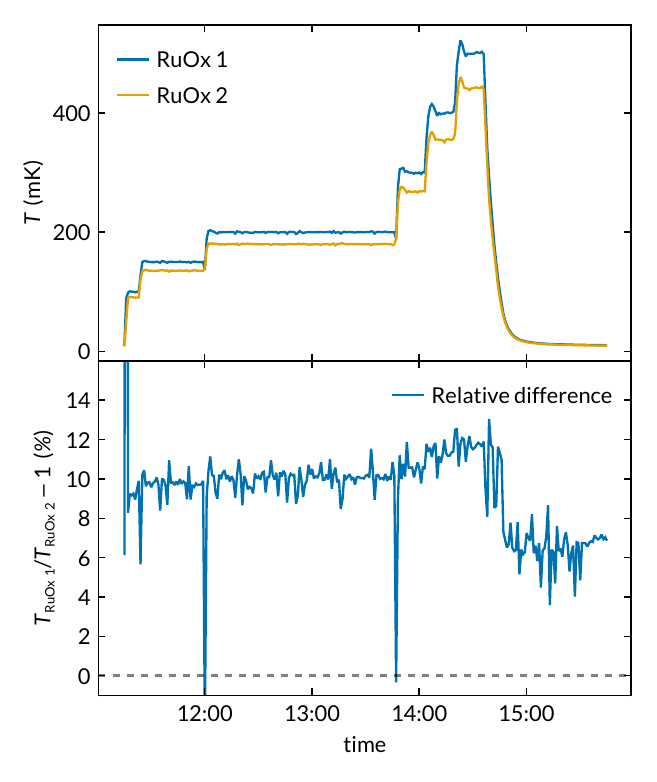}
    \caption{The temperature of two RuOx thermometers mounted to the MXC plate of the cryostat over a range of temperatures. In the presented experiments, RuOx 1 was used. There is a nearly constant offset between the two sensors, suggesting a poor calibration or drift due to aging. The experiment was repeated with the position of RuOx 2 changed, which resulted in a similar deviation. This implies that the MXC plate is uniformly thermalized. }
    \label{fig:ruox-difference.}
\end{figure}

\end{document}